# Strong phonon-coupled trion emission in Janus homobilayer

Tianyishan Sun[1,2], Dingkun Bi[1,2], Haonan Wang[3], Hiroto Ogura[1,2], Yanlin Gao[4,6], Mina Maruyama[5,6], Susumu Okada[4,6], Shiro Entani[7], Kenji Watanabe[8], Takashi Taniguchi[9], Kazunari Matsuda[3], and Toshiaki Kato[1,2]

[1] Graduate School of Engineering, Tohoku University, Sendai 980-8579, Japan

[2] Advanced Institute for Materials Research (AIMR), Tohoku University, Sendai 980-8577, Japan

[3] Institute of Advanced Energy, Kyoto University, Uji, Kyoto 611-0011, Japan

[4] Department of Physics, Graduate School of Pure and Applied Sciences, University of Tsukuba, Tsukuba 305-8571, Japan

[5] Department of Materials Science, Graduate School of Pure and Applied Sciences, University of Tsukuba, Tsukuba 305-8571, Japan

[6] Tsukuba Institute for Advanced Research (TIAR), University of Tsukuba, Tsukuba 305-8577, Japan

[7] Quantum Materials and Applications Research Center, National Institutes for Quantum Science and Technology (QST), Takasaki 370-1292, Japan

[8] Research Center for Electronic and Optical Materials, National Institute for Materials Science (NIMS), Tsukuba 305-0044, Japan

[9] Research Center for Materials Nanoarchitectonics, National Institute for Materials Science (NIMS), Tsukuba 305-0044, Japan

# Abstract

Janus transition metal dichalcogenides with broken mirror symmetry provide a novel platform for exploring many-body effects mediated by spin-valley-phonon interactions. Strong electron-phonon coupling demands a precise match between a strong effective dipole moment and nanometer-scale charge confinement. Here, we report thermally robust and enhanced trion emission in a symmetric bilayer WSeS with an S/S stacking configuration. The inward-pointing antiparallel built-in electric fields induce a strong electron confinement effect at the interface, driving dominant trion emission at cryogenic temperatures. Furthermore, a massive electron-phonon coupling strength ($\approx 3.9$) resonant with the optical phonon $A_1^2$ energy ($\approx 51$ meV) contributes to the thermal stability of this trion emission at elevated temperatures. By tailoring the interfacial architecture in Janus WSeS homobilayers, this study overcomes the thermal decoherence limits inherent to traditional semiconductor systems, offering novel insights into room-temperature many-body interactions.

# Introduction

Monolayer transition metal dichalcogenides (TMDs) have emerged as a versatile two-dimensional (2D) platform for next-generation optoelectronic and quantum devices [1,2]. The broken in-plane inversion symmetry in the 1H phase gives rise to coupled spin and valley degrees of freedom [3], driving extensive research into valleytronics [4,5]. Beyond these electronic properties, there is growing interest in introducing the phonon degree of freedom into TMDs. Coupling phonons with spin and valley dynamics could unlock novel functionalities, paving the way for advanced spin-valley-phonon hybridized devices [6].

Introducing the phonon degree of freedom requires substantially enhancing electron-phonon interactions, which are dominated by the Fröhlich mechanism in polar semiconductor monolayers [7]. The Fröhlich coupling strength is governed by two key factors: the magnitude of the polar lattice field (i.e., effective dipole moment) and the characteristic scale of spatial charge confinement [7,8] (**Fig. 1a**). According to exciton-phonon coupling theory in low-dimensional systems, Fröhlich interaction is maximized in the intermediate confinement regime, where the spatial confinement scale approaches the intrinsic exciton Bohr radius [9]. At this characteristic scale, spatial mismatch between electron and hole wavefunctions is optimized, thereby maximizing the macroscopic polarization field [9,10]. In monolayer TMDs, the exciton Bohr radius is theoretically estimated [11] and experimentally verified to be approximately 1 nm [12,13], offering an ideal spatial dimension for strong exciton-phonon coupling.

To date, exciton-phonon interactions in TMDs have been explored primarily in defect-bound excitons [14,15] and moiré superlattices [16], both of which rely on localized trapping potentials or in-plane polarizations. In moiré superlattices (at a given potential depth), the confinement scale can be tuned by the twist angle and is 5–10 nm [17], whereas trapping at point defects is uncontrollable and excessively localized (<1 nm) [18]. However, the in-plane polarization field induced by moiré superlattices is near zero [19]. By contrast, bilayer TMDs exhibit an electron confinement scale near 1 nm. Nevertheless, the out-of-plane dipole moment in a TMD heterolayer is weak because it arises from the spatial separation of electrons and holes in different TMD monolayers. To overcome these structural limitations, we focused on Janus TMDs [20,21]. Due to broken out-of-plane mirror symmetry, Janus monolayers possess built-in vertical dipole moments [20-22]. Therefore, stacking Janus monolayers can generate intense interlayer confinement on a 1 nm scale along the same out-of-plane axis as the intrinsic dipole moment (Fig. 1b), thereby precisely targeting the theoretical sweet spot for Fröhlich coupling. Thus, Janus homobilayer architectures represent an extraordinarily promising platform for engineering strong electron-phonon interactions, bridging the spatial gap between defect- and moiré-scale physics.

In this study, we demonstrate that S/S-stacked Janus WSeS homobilayers exhibit a thermally robust trion emission characterized by exceptionally strong optical phonon coupling. Temperature-dependent photoluminescence (PL) spectroscopy reveals a colossal dimensionless phonon-coupling constant (≈ 3.9), vastly exceeding values reported for traditional quantum wells and TMD moiré superlattices. Furthermore, this strong coupling is mediated by a high-energy optical phonon (≈ 51 meV), which significantly exceeds the room-temperature thermal energy limit ($k_B T \approx 26$ meV, where $k_B$ is the Boltzmann constant and $T$ is the absolute temperature). Cryogenic Raman spectroscopy confirms that this effective phonon energy resonates almost perfectly with the out-of-plane $A_1^2$ optical phonon of Janus WSeS, establishing the robust optical feature as a polaronic trion state. Our findings demonstrate Janus TMD homobilayers as a highly effective architecture for exploring room-temperature strong phonon-coupling physics and novel many-body quantum states.

# Results

## Excitonic properties of Janus WSeS homobilayers

**Fig. 1c** displays the room-temperature PL spectra of three WSeS bilayer configurations, with the monolayer (1L) serving as a reference. The spectral profiles were deconvoluted using Voigt functions to identify distinct excitonic complexes [23]. The red-shaded component is assigned to neutral exciton emission ($X^0$). Compared with monolayer WSeS, $X^0$ in the bilayers exhibits attenuated intensity and spectral broadening, a trend consistent with the direct-to-indirect bandgap transition in conventional TMD homobilayers [1,24]. The variations in the absolute emission energies among samples stem from the Janus conversion process, which introduces local fluctuations in dielectric environments, built-in strain, and the degree of sulfurization.

Beyond the neutral exciton, distinct lower-energy features (blue-shaded) are observed across all configurations. Because the characteristic trion binding energy in standard TMDs typically ranges from 30 to 40 meV [25-27], the lower-energy emissions in the 1L, Se/S, and S/S configurations (with energy separations ($\Delta E$) of 33.0 meV, 40.9 meV, and 40.0 meV, respectively) are unambiguously assigned to trion states (T). The disparity in the trion-to-exciton intensity ratio between the Se/S and S/S bilayers is attributed to the distinct localized interfacial potentials induced by different chalcogen arrangements (**Fig. S1**). In contrast, the Se/Se-ordered bilayer displays a substantially larger energy separation of 93.7 meV, ruling out a trion origin and instead pointing to indirect excitons (IX) or localized excitons (LX). These room-temperature PL results demonstrate that the intrinsic built-in electric field profoundly modulates the optical emission characteristics of Janus WSeS homobilayers.

Full hexagonal boron nitride (hBN) encapsulation suppresses inhomogeneous broadening [28], enabling precise identification of excitonic species in the cryogenic PL spectra at 4.5 K (**Fig. 1d**). These peak assignments are further supported by power-dependent PL measurements (**Fig. S2**), where features corresponding to neutral excitons and trion complexes exhibit linear power scaling. Among the three stacked configurations, the Se/S-ordered homobilayer displays a spectral profile similar to that of the monolayer, with more pronounced localized exciton emission at lower energies. Conversely, the S/S-ordered bilayer exhibits complete quenching of the neutral exciton alongside remarkably intense trion emission. This dominant trion emission is consistently reproduced across three additional S/S-ordered bilayer samples (**Fig. S3**).

This cryogenic behavior corroborates the room-temperature observations, where the S/S stacking yields the highest trion-to-exciton intensity ratio (**Fig. S1**). Such selective and dominant trion formation underscores a strongly enhanced charge accumulation and transfer mechanism specific to the S/S interface, establishing it as a highly promising platform for investigating many-body physics.

## Electrostatic doping and gate-dependent PL behavior

To elucidate the origin of the dominant trion emission in the S/S-ordered bilayer, electrostatic doping experiments were performed. **Fig. 2a,b** show the schematic illustration and optical microscopy image of the hBN-encapsulated S/S bilayer device. The structure employs a top few-layer graphene electrode to apply a vertical electric field, with the bottom graphene layer acting as a grounded contact. We first investigated the gate-dependent PL response of monolayer WSeS (**Fig. S4**) as a baseline. As depicted in **Fig. 2c,d**, sweeping the gate bias yields a characteristic electrostatic evolution, where the higher-energy neutral exciton $X^0$ dominates the spectrum at positive gate bias. Sweeping the bias toward negative values leads to a clear spectral transition, characterized by the quenching of $X^0$ and the emergence of two lower-energy negative trion complexes. This spectral evolution aligns well with previously reported gate-dependent reflectance contrast spectra of monolayer WSeS [29]. The two trion complexes, separated by an energy difference of about 13 meV, are assigned to intravalley ($T_{intra}$) and intervalley ($T_{inter}$) trions, respectively.

In contrast to the highly tunable monolayer, the S/S-ordered bilayer displays a remarkably robust spectral profile across the gate voltage range from -10 V to +10 V (**Fig. 2e,f**). The spectrum at 0 V exhibits excellent consistency, reproducing the dominant trion emission observed in ungated samples (**Fig. S3**). Driving the gate bias to negative values to accumulate additional electrons enhances the trion peak intensity alongside a subtle redshift. This intensity enhancement driven by the elevated electron density confirms it as a negatively charged trion. A similar gate-induced redshift has been observed in monolayer $WSe_2$ [30], monolayer WSeS [29], and bilayer $WS_2$ [31]. This effect is attributed to an enhanced trion binding energy caused by the quantum-confined Stark effect [30,31], induced by the strong out-of-plane electric field applied across the S/S bilayer. However, even when sweeping the gate bias to +10 V to deplete electrons, the neutral exciton $X^0$ remains absent. Although the primary trion peak significantly diminishes in intensity, a broad, lower-energy emission band originating from localized defect states emerges. Consistent gate-dependent PL behavior was also confirmed in a second device (**Fig. S5**).

The inability to recover neutral exciton emission $X^0$ even under high positive gate bias indicates that the S/S-ordered bilayer is intrinsically locked in a heavily n-doped state. This behavior is attributed to the unique face-to-face arrangement of the sulfur layers, wherein the opposing out-of-plane built-in electric fields of the two constituent Janus monolayers form a deep localized potential well at the interface. This interfacial potential well serves as a strong electron confinement, pinning the Fermi level deep within the conduction band and rendering the optical response largely immune to electrostatic tuning.

## Enhanced electron-phonon coupling in S/S-ordered bilayers

To clarify the excitonic dynamics and the microscopic origin of the dominant trion emission in S/S-ordered bilayer WSeS, temperature-dependent PL measurements were performed from 4.5 K to 300 K (**Fig. 3a and S5**). The trion emission is primarily composed of two distinct features, $T_1$ and $T_2$, which evolve differently as a function of increasing temperature. $T_1$ dominates at cryogenic temperatures but rapidly quenches with increasing temperature, whereas $T_2$ becomes increasingly prominent at elevated temperatures and persists up to room temperature. Neutral exciton emission remains heavily suppressed below 80 K. The robust thermal stability of trions in the S/S bilayer is further corroborated by its remarkably high thermal activation energy ($E_a \approx 33.6$ meV) for trion dissociation, which represents the highest value among monolayer and other bilayer WSeS configurations (**Fig. S6**).

Power-dependent PL measurements at 4.5 K reveal distinct scaling behaviors for $T_1$ and $T_2$ (**Fig. 3c**). $T_1$ is assigned to a superposition of localized and free trions, as evidenced by its crossover from sublinear growth ($\alpha_{T1_1} \approx 0.7$) at low excitation densities to linear growth ($\alpha_{T1_2} \approx 1.2$) under higher powers. The bound components within $T_1$ likely stem from the deep localized potential at the S/S interface due to the chalcogen vacancies. In contrast, $T_2$ displays a strict linear power dependence ($\alpha_{T2} \approx 1.2$) across the entire power range, characteristic of a free trion state [32].

To quantify exciton-phonon interactions, the temperature-dependent energy shifts of $X^0$ and $T_2$ were fitted using the O'Donnell model (**Equ. 1**) [33], as depicted in **Fig. 3b**.

$$E_g(T) = E_g(0) - S\langle\hbar\omega\rangle\left(\coth\frac{\langle\hbar\omega\rangle}{2k_B T} - 1\right) \qquad \text{Equation 1}$$

$E_g$ (0) is the transition energy at 0 K, $S$ is the dimensionless exciton-phonon coupling constant, and $\hbar\omega$ represents the average effective phonon energy. Compared to other configurations [34,35] (**Fig. 3d and S7**), $T_2$ in the S/S bilayer yields a remarkably high phonon coupling strength ($S \approx 3.9$) and a large effective phonon energy ($\hbar\omega \approx 51.0$ meV). This elevated $S$ parameter signifies exceptionally strong electron-phonon coupling associated with the $T_2$ state [36].

To identify the vibrational modes driving this coupling, cryogenic Raman spectroscopy at 4.5 K was conducted (**Fig. 3e and S8)**. Converting the Raman shift to an energy scale reveals that the effective phonon energy ($\hbar\omega \approx 51$ meV) of $T_2$ extracted from O'Donnell fitting aligns almost perfectly with the out-of-plane optical phonon mode $A_1^2$ at about 51 meV ($\approx$ 411 $cm^{-1}$) in Janus WSeS [37]. This precise energetic resonance and high $S$ value directly evidence that photogenerated charges interact strongly with out-of-plane lattice vibrations. Furthermore, compared to other WSeS materials (**Fig. S8**), an emergent out-of-plane phonon mode appears uniquely in the S/S configuration at around 52 meV, which can be attributed to severe interfacial strain and strong electrostatic repulsion between opposing sulfur atom planes. The enhanced peak intensity and broadened linewidth of this mode further confirm an elevated electron-phonon coupling strength for the $T_2$ state [38]. These amplified out-of-plane vibrations provide an efficient scattering channel, thereby characterizing the emission as a heavily $A_1^2$ phonon-coupled trion. Consequently, this robust emission feature is best conceptualized as a polaronic trion mediated by out-of-plane optical phonons [39].

To probe the formation kinetics of this trion, time-resolved photoluminescence (TRPL) measurements were conducted (**Fig. 3f**). Because $T_1$ dominates at 4.5 K, the TRPL response captures the integrated dynamics of $T_1$ and $T_2$. Notably, the PL rise time of the trion in the S/S bilayer is as fast as the neutral exciton in monolayer WSeS, nearing the instrument response function (IRF) limit. Standard trion formation in monolayer TMDs typically exhibits a finite delay, corresponding to the time required for a photogenerated exciton to capture a free carrier [40,41]. The decay lifetime is extracted by the deconvolution of a biexponential function as $I(t) = A_1\exp(\tau_1/t) + A_2\exp(\tau_2/t)$, where $A_1$, $A_2$, and $\tau_1$, $\tau_2$ are the amplitudes and decay lifetimes. The total lifetime for each complex can be explained as $\tau = (A_1\tau_1+A_2\tau_2)/(A_1+A_2)$. Therefore, the shortest lifetime and near-instantaneous rise indicate strong phonon-coupled trion emission in the S/S bilayer.

Furthermore, this ultrafast, heavily phonon-coupled trion formation effectively suppresses intervalley scattering, as verified by circular polarization PL measurements (**Fig. S9**). Despite its heavy effective mass and complex structure, the trion in the S/S bilayer yields a degree of circular polarization (DOCP) of approximately 0.24, surpassing the value of 0.18 measured in monolayer WSeS. The DOCP is described as DOCP = $P_0/(1+\tau/\tau_v)$, where $P_0$ is the initial valley polarization, $\tau$ is the total excitonic decay lifetime, and $\tau_v$ is the intervalley depolarization time. Assuming $P_0$ and $\tau_v$ do not show strong fluctuations with stacking orders, the shortest lifetime $\tau$ of the trion in S/S configurations is consistent with its highest DOCP. The superior preservation of valley polarization and ultrafast decay lifetime in S/S bilayer points to a distinct kinetic pathway: photogenerated holes are rapidly captured by the strong interfacial potential well, driving ultrafast trion formation without kinetic bottlenecks. The detailed mechanism is introduced in the following section.

## DFT analysis for electronic structure of homobilayer WSeS

Based on the electronic properties of the Janus homobilayer of WSeS using density functional theory (DFT) calculations of the electronic wavefunctions and localized potentials (**Fig. 4a,b and S10**), we will discuss the physical origin of the optical phonon-coupled trion. In the Se/S configuration with parallel built-in electric fields, the wavefunction is asymmetrically skewed toward one layer. In the Se/Se stacking with antiparallel electric fields pointing outward, the wavefunctions of the valence and conduction band edges are symmetrically distributed except for the conduction band edge at the K point (CB@K), owing to the small splitting of the conduction band edges. Conversely, wavefunctions of the band edges of a Janus homobilayer with the S/S interface exhibit symmetric distributions across the layers. This symmetric distribution is driven by the face-to-face arrangement of the highly electronegative sulfur atoms. As shown in the electrostatic potential profile across the layers (**Fig. 4b**), the opposing Janus built-in electric fields create a deep and narrow electrostatic positive potential well, i.e., a hole trap between the layers. The valley on the sulfur side is deeper than that on the selenium side, clearly corroborating that the strong Janus built-in electric field is directed from the outside toward the inside of the interface.

Based on these experimental and theoretical insights, we formulate a microscopic model to account for the ultrafast and dominant trion emission in S/S-ordered homobilayers (**Fig. 4c**). First, opposing Janus electric fields efficiently trap photogenerated holes within an ultrathin (≈ 1 nm) region at the S/S interface. Second, owing to the intrinsic heavy n-type background, the trapped hole serves as a localized Coulombic attraction core, rapidly binding surrounding free electrons. This charge accumulation mechanism explains the IRF-limited TRPL rise time (**Fig. 3f**) and the preserved circular polarization of the trion emission in the S/S bilayer. Finally, the extreme spatial confinement of these charges within a near 1 nm interfacial scale matches the optimal localization length for Fröhlich-type electron-phonon interactions. The interplay between nanoscale charge confinement and strong localized lattice deformation gives rise to the exceptionally large exciton-phonon coupling constant ($S \approx 3.9$), establishing the absolute dominance of trion emission across both wide temperature and electrostatic gating ranges.

## Modulation of the trion emission via dielectric spacers

To explore the tunability of the trion emission in S/S-ordered architecture, homobilayers with hBN spacers of varying thicknesses inserted between the Janus monolayers were investigated (**Fig. 5a,b**). Excitation power-dependent PL measurements confirm the free excitonic nature of all primary emission features (**Fig. S2**). When a few-layer hBN spacer is inserted, the two Janus monolayers become electrostatically decoupled. Consequently, the temperature-dependent PL spectra (**Fig. 5c**) mirror typical monolayer WSeS behavior (**Fig. S6**), where both neutral exciton and trions coexist at low temperatures, and trion emission rapidly quenches above 120 K. In contrast, inserting a monolayer hBN spacer preserves the anomalous excitonic dynamics observed in the unspaced S/S-ordered bilayer. As depicted in **Fig. 5d**, trion emission remains dominant over the neutral exciton up to room temperature. Because the 1L hBN is sub-nanometer in thickness, it is electrostatically transparent, allowing the opposing Janus built-in electric fields to penetrate the spacer and preserve the deep interfacial potential well required for ultrafast hole trapping.

Exciton-phonon coupling parameters for the free trion $T_2$ in the S/1L hBN/S and S/few-layer hBN/S configurations were extracted via the O'Donnell model (**Fig. S11**), yielding a phonon coupling constant $S \approx 2.0$ and an effective phonon energy $\hbar\omega \approx 12.6$ meV for 1L hBN separated, as well as much weaker $S \approx 1.7$ and $\hbar\omega \approx 12.6$ meV for few-layer hBN-separated S/S bilayer WSeS, respectively. These values are comparable to those measured in monolayer WSeS (**Fig. S7**), demonstrating that physical separation by hBN suppresses direct interfacial optical phonon scattering. Therefore, the electron confinement scale can be modulated by inserting an hBN spacer, leading to a reduced phonon coupling strength (**Fig. 1a**). Nevertheless, the dominant and thermally robust trion emission in the 1L hBN-separated S/S bilayer WSeS is attributed to the antiparallel dipole moment across individual monolayers.

The thermal stability of trions in S/S configurations is quantitatively demonstrated by the integrated trion-to-exciton intensity ratio ($I_T/I_{X0}$) as a function of temperature (**Fig. 5e**). At 4.5 K, both unspaced and 1L hBN-spaced S/S configurations exhibit $I_T/I_{X0}$ ratios exceeding those of the reference samples (Se/S, 1L WSeS, and S/few-layer hBN/S) by over an order of magnitude. This ratio remains strictly greater than 1 up to 240 K for both S/S architectures, confirming that both deep interfacial electrostatic trapping and strong phonon dressing stabilize trion states against thermal dissociation.

Furthermore, **Fig. 5f** benchmarks the exciton-phonon coupling strength $S$ against the effective phonon energy $\hbar\omega$ across representative quantum materials [18, 42-48]. For polaronic trion states to survive at room temperature, the interacting optical phonon energy must exceed the thermal energy threshold ($k_B T \approx 26$ meV) [49]. While conventional quantum wells exhibit high phonon energies ($\hbar\omega \approx 90$ meV), they suffer from weak coupling strengths ($S < 1$) due to soft lattice interactions [42-44]. Conversely, standard TMDs [34,45,46] and moiré superlattices [47] offer moderate coupling strengths but remain constrained by phonon energies comparable to or below $k_B T$. The S/S-stacked Janus WSeS homobilayer overcomes these trade-offs by combining deep nanoscale electrostatic confinement with strong local out-of-plane dipole interactions. This interfacial design achieves an exceptionally large coupling strength alongside a high optical phonon energy well above the room-temperature thermal energy limit.

## Discussion

In summary, S/S-ordered Janus WSeS homobilayers host enhanced and thermally robust trions governed by a synergistic combination of intense dipolar electrostatic trapping ($\approx 1$ nm) and colossal electron-phonon coupling ($S \approx 3.9$, $\hbar\omega \approx 51$ meV). This intrinsic dipolar architecture accelerates ultrafast trion formation, suppresses intervalley scattering, and prevents thermal dissociation up to room temperature. Interfacial engineering with a 1L hBN allows independent tuning of phonon coupling strength via extending the confinement scale. By overcoming the thermal constraints of standard semiconductors, this platform opens promising avenues for uncooled room-temperature valleytronics, polaronic charge transport, and spin-valley-phonon quantum transducers.

# Materials and Methods

Monolayer $WSe_2$, bilayer $WSe_2$, few-layer hBN, and few-layer graphene were mechanically exfoliated by blue tape (Nitto) from the bulk crystals (2D Semiconductors for $WSe_2$, NIMS for hBN, and HOPG for graphene) onto 280-nm $SiO_2$/Si substrates. Monolayer hBN flakes were fabricated through ion irradiation [50]. Obtained monolayer and bilayer $WSe_2$ were confirmed by the contrast of the optical microscope images, Raman, and PL spectroscopies.

**Janus treatment**

Replacement of top selenium with sulfur in $WSe_2$ was achieved by a plasma-assisted room-temperature sulfurization process [51,52]. The starting materials were placed in a 2 cm-wide quartz tube, which was then evacuated to lower than 0.1 Pa. 99.9% pure hydrogen ($H_2$) gas with a flow rate of 20 sccm was filled in to maintain the pressure at about 60 Pa. The $H_2$ plasma was ignited by a copper coil with 30 W RF power. The sample and sulfur powder (98%, Nilaco) were placed downstream and upstream of the $H_2$ flow, and far away from the copper coil by 1 cm, respectively. The total time of the sulfurization was about 12 ~ 20 min, depending on the starting materials.

**Fabrication of the hBN-encapsulated Janus homobilayers**

Se/S: monolayer $WSe_2$ and few-layer hBN were picked up sequentially by a polymer (Acrylic resin)-assisted-dry transfer process [53] and then released onto a clean 280-nm $SiO_2$/Si substrate. After sulfurizing the $WSe_2$/hBN stack, another monolayer $WSe_2$ was transferred onto this WSeS/hBN, followed by the second sulfurization. The top few-layer hBN was finally transferred onto the SWSe/SWSe/hBN stack to protect the materials.

S/S: two starting materials were used for this sample. One is also the $WSe_2$/hBN stack, and another is the $WSe_2$/hBN on the polymer used for dry transfer. After sulfurizing these two materials, the SWSe/hBN on the polymer was aligned and released onto the bottom SWSe/hBN stack to get the hBN-encapsulated SeWS/SWSe.

Se/Se: the starting material was an exfoliated bilayer $WSe_2$ on the normal $SiO_2$/Si substrate. After replacing the top Se atoms, the SWSe/SeWSe was picked up by hBN on polymer. The stack on the polymer was sulfurized second to get the SWSe/SeWS/hBN. Finally, this stack was released onto a clean hBN.

All obtained bilayer samples were thermally annealed at 200 °C under 0.1 Pa for 2 hours to enhance interlayer coupling.

**Optical measurements**

Power-dependent PL, temperature-dependent PL, and Raman measurements were carried out using a Horiba LabRAM *HR* spectrometer. Samples were excited by a continuous-wave (CW) laser with a wavelength of 532 nm (2.33 eV), focused onto the target area through a 50× objective lens with a numerical aperture of 0.5. For spectral dispersion, gratings of 100 grooves/mm and 2400 grooves/mm were employed for PL and Raman collection, respectively. Low-temperature (4.5 K) optical measurements were performed in an Oxford Instruments Microstat He. The temperature was precisely monitored and regulated by a Mercury iTC cryogenic environment controller from Oxford Instruments.

A pulsed supercontinuum light source with a repetition rate of 5 MHz and a center photon energy of 2.17 eV was employed for the time-resolved PL spectroscopy. A 50× objective lens with a numerical aperture of 0.67 was used for focusing the excitation laser light and acquiring optical images. The samples were positioned in a cryogen-free cryostat whose temperature ranged from 5.5 K to room temperature. The emission signals were transported through an optical fiber and detected via a Si avalanche photodiode using a time-correlated single-photon counting technique.

For the circular polarization-resolved PL spectra measurement, a linearly polarized CW laser (1.95 eV) was employed as the excitation source. The polarization results of the emission signal were determined by measuring PL spectra at various detection angles, which were rotated via a combination of a half-wave plate and a linear polarizer. The emission signals were coupled with an optical fiber and detected using a spectrometer and charge-coupled device.

**Computational methods**

DFT calculations are carried out using the STATE program package. The exchange-correlation potential for interacting electrons is treated using vdW-DF2 with C09 functional, which includes weak dispersive interactions in atomic-layer materials. Ultrasoft pseudopotentials are used to express the interaction between valence electrons and ions. The valence wave function and deficit charge density are expanded by plane-wave basis sets with the cutoff energies of 25 and 225 Ry, respectively. The Brillouin zone integration was carried out 20 × 20 k-meshes. All atoms were fully optimized until the force acting on each atom was less than 0.005 Ry/Å. To avoid unphysical dipole interactions with the imaging cells, we used the effective screening medium method, which allows us to set an open boundary condition normal to the layers in terms of the electrostatic potential [54].

## Acknowledgment

Special thanks go to Louis Smet for his insightful discussions and comments about data analysis. This research was supported by JSPS KAKENHI (Grant Nos. P22H05441, JP23H00097, JP23K17756, JP23K19193, JP24H01165, JP24K17584 and JP26K22456), JST-CREST (Grant No. JPMJCR23A2), JST SPRING (Grant No. JPMJSP2114), KONDO-ZAIDAN, MATSUDA-ZAIDAN, and the Murata Science Foundation, and the Cooperative Research Project Program of the Research Institute of Electrical Communication, Tohoku University.

## Contributions

T.K. supervised and designed this project. T.S., T.K., and H.O. contributed to the writing, revision, and discussion of this manuscript. T.S. contributed to the preparation and basic characterization of Janus TMD bilayers. T.S. and D.B. contributed to the fabrication of gate-controlled samples and the cryogenic optical measurements. H.W. and K.M. carried out the TRPL and circular polarization PL measurements. Y.G., M.M., and S.O. contributed to the DFT calculations. S.E. provided the synthesized monolayer hBN. K.W. and T.T. provided hBN samples.

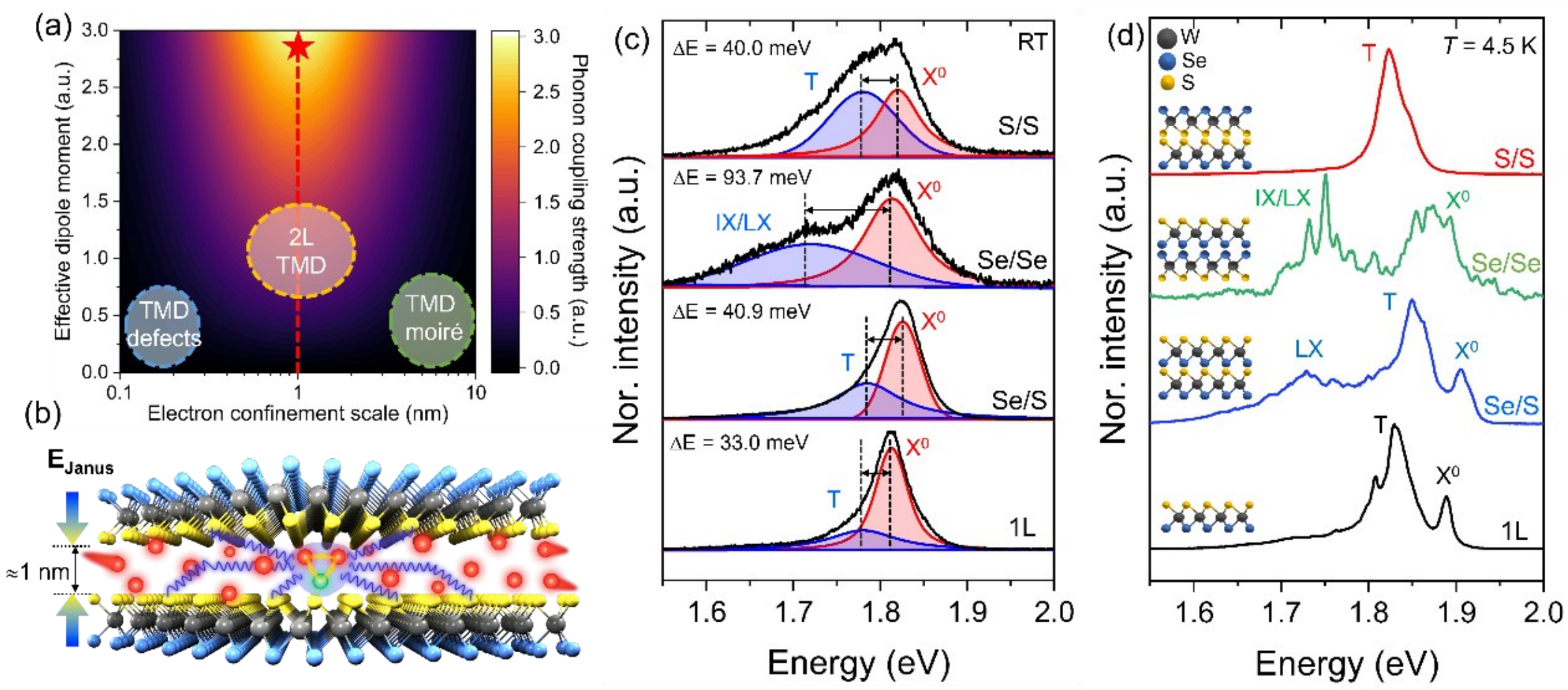

**Fig. 1**: **Optical properties in Janus WSeS homobilayers.** (a) Conceptual contour map illustrating the qualitative interplay between the electron confinement scale, effective dipole moment, and phonon coupling strength [7,8]. (b) Schematic illustration of the electron confinement effect induced by the Janus out-of-plane electric field ($E_{Janus}$) in a symmetric homobilayer Janus TMD. Wavelines represent the phonon scattering of the trion emission. Blue, grey, yellow, red, and green spheres are selenium, tungsten, and sulfur atoms, as well as electrons and holes. (c, d) PL spectra of monolayer and homobilayer WSeS across three distinct stacking configurations at (c) room temperature and (d) 4.5 K. Red- and blue-shaded peaks in (c) represent deconvoluted Voigt components corresponding to different excitonic species. Insets in (d) show the corresponding atomic structures.

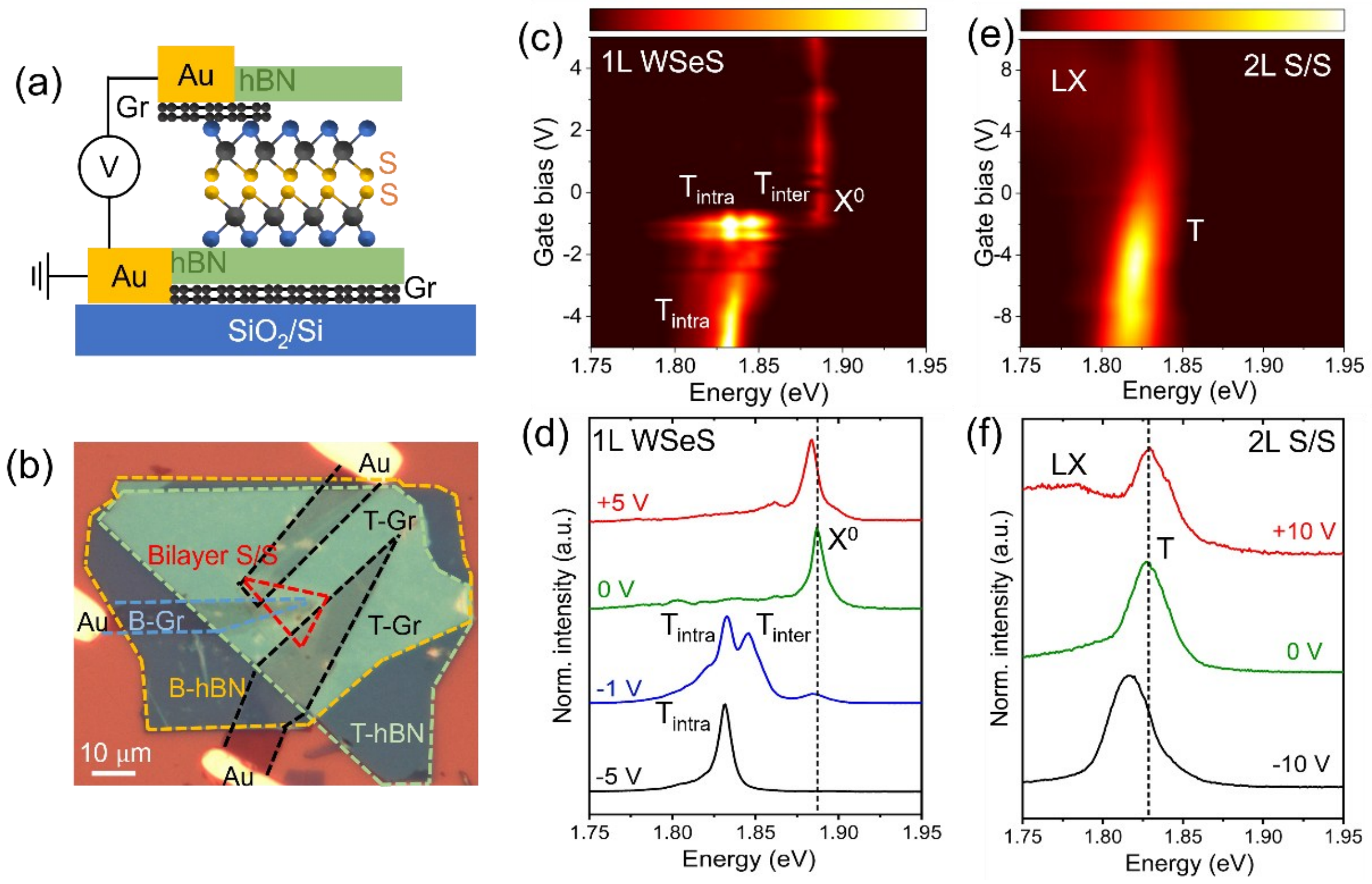

**Fig. 2**: **Gate dependence of PL spectra in monolayer and bilayer WSeS with S/S configuration.** (a) Schematic diagram of the hBN-encapsulated S/S-ordered WSeS homobilayer device with top and bottom few-layer graphene electrodes. (b) Optical microscopy image of the device. (c) 2D color mapping and (d) selected line cuts of gate-dependent PL spectra for monolayer WSeS at 4.5 K. (e) 2D color mapping and (f) selected line cuts of gate-dependent PL spectra for the S/S-ordered WSeS homobilayer at 4.5 K. Each excitonic component is labelled in (c)-(f).

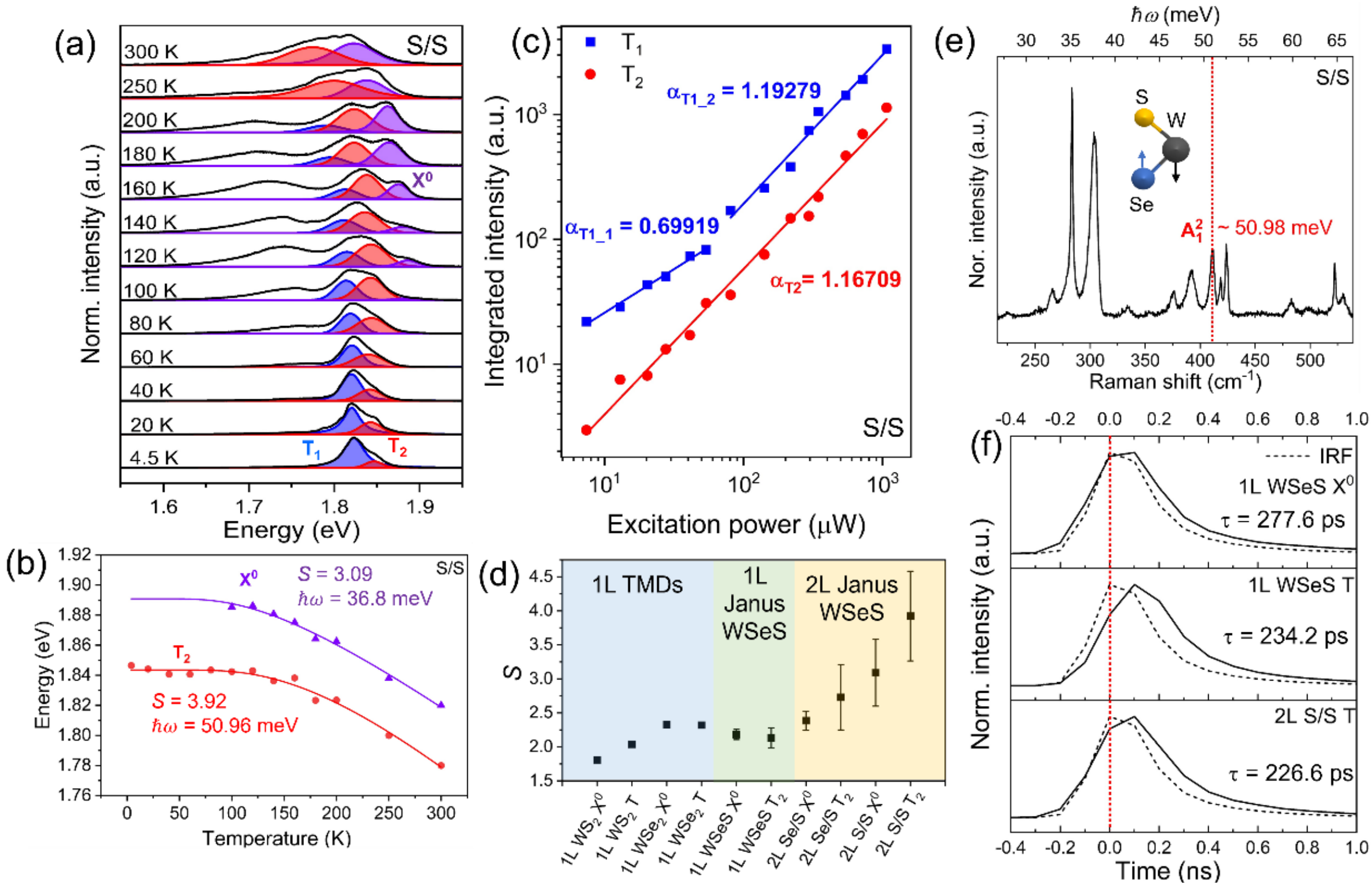

**Fig. 3**: **Phonon-coupled trion emission in S/S-ordered WSeS bilayer.** (a) Temperature evolution of the PL spectra of the S/S-ordered WSeS homobilayer from 4.5 K to 300 K. The blue-, red-, and purple-shaded peaks represent two different trion states, $T_1$ and $T_2$, and the neutral exciton $X^0$. (b) Temperature dependence of peak energies for $T_2$ and $X^0$, fitted with the O'Donnell model (solid lines). (c) Integrated PL intensities of $T_1$ and $T_2$ as a function of excitation power at 4.5 K. (d) Comparison of exciton-phonon coupling constants (S) extracted from the O'Donnell model across monolayer $WS_2$ [34], monolayer $WSe_2$ [35], monolayer WSeS, and various WSeS homobilayer configurations. (e) Cryogenic Raman spectrum of the S/S-ordered bilayer at 4.5K. The out-of-plane optical phonon mode $A_1^2$ at 50.98 meV is highlighted with a red dashed line and illustrated in the inset. (f) TRPL decay profiles of the neutral exciton and trion in monolayer WSeS, alongside the trion in the S/S-ordered bilayer. Dashed lines represent the instrument response function. The total decay lifetime $\tau$ is listed for each trace.

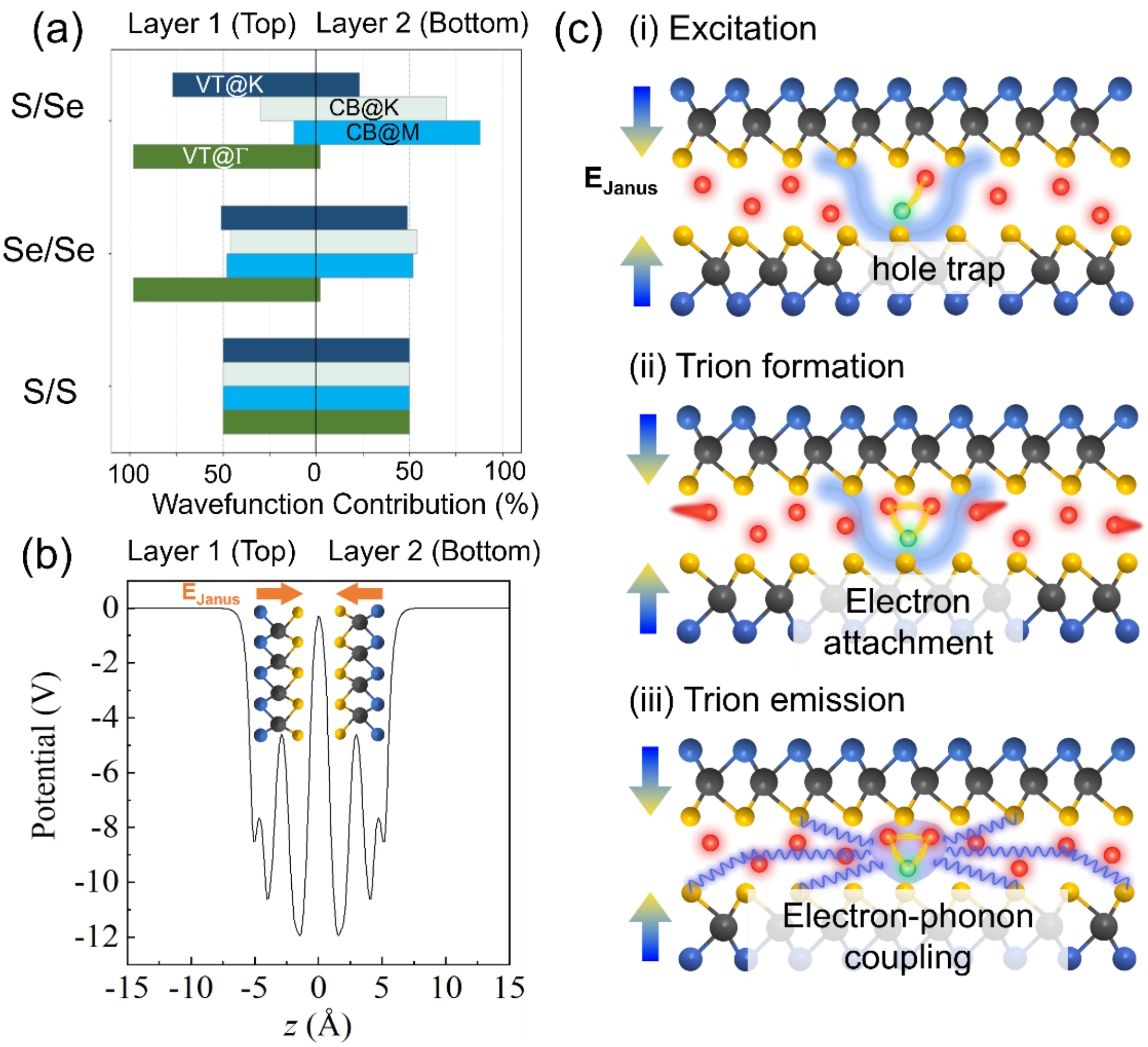

**Fig. 4**: **Electronic structure of Janus WSeS homobilayers and the mechanism of trion formation in S/S-ordered bilayer.** (a) Distribution ratio of wavefunctions of the highest valence band at K (VT@K) and at Γ (VT@Γ) and those of the lowest conduction band at K (CB@K) and M (CB@M). (b) The plane-averaged electrostatic potential of the S/S-stacked WSeS homobilayer. The gray, yellow, and blue spheres represent W, S, and Se atoms, respectively. (c) Schematic diagram illustrating the microscopic mechanism of ultrafast hole trapping, electron attachment, and strong optical phonon coupling during negative trion formation in the S/S-ordered WSeS homobilayer.

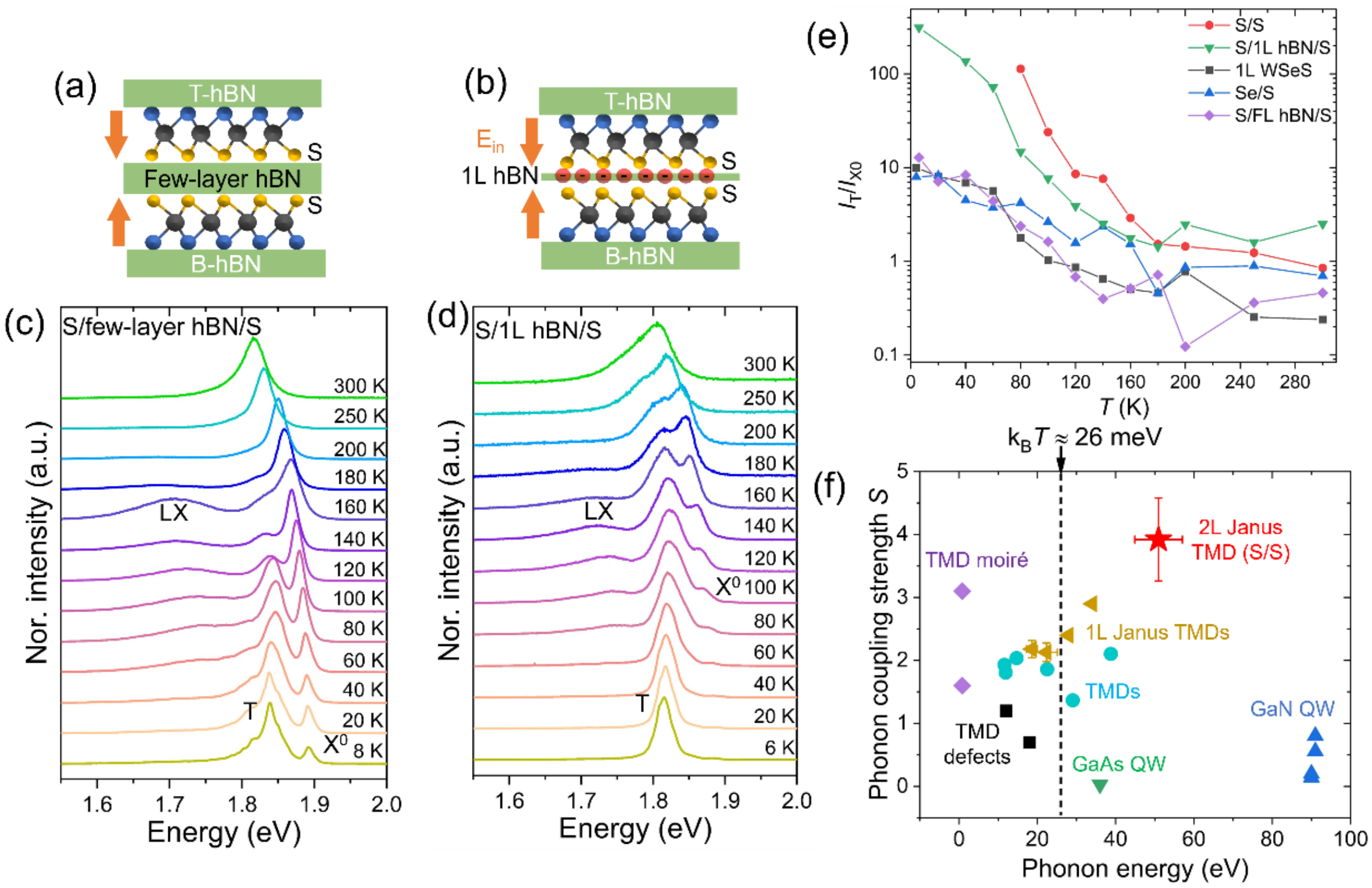

**Fig. 5**: **Interfacial engineering by dielectric spacers.** (a, b) Schematic diagrams of S/S-ordered WSeS homobilayers separated by (a) few-layer and (b) monolayer hBN spacers. (c, d) Temperature-dependent PL spectra of (c) S/few-layer hBN/S and (d) S/1L hBN/S configurations from cryogenic temperatures to 300 K. (e) Integrated trion-to-neutral-exciton intensity ratio ($I_T/I_{X0}$) as a function of temperature across various Janus configurations and monolayer controls. (f) Benchmark of the electron-phonon coupling strength S to the average phonon energy $\hbar\omega$ across diverse quantum material platforms, including GaN quantum well (QW) [42,43], GaAs QW [44], TMDs [34, 45, 46], TMDs moiré [47], TMD defects [18], and monolayer Janus TMDs [48]. The S/S-ordered Janus WSeS homobilayer in this work is highlighted with a red star.

## Supplementary Figures

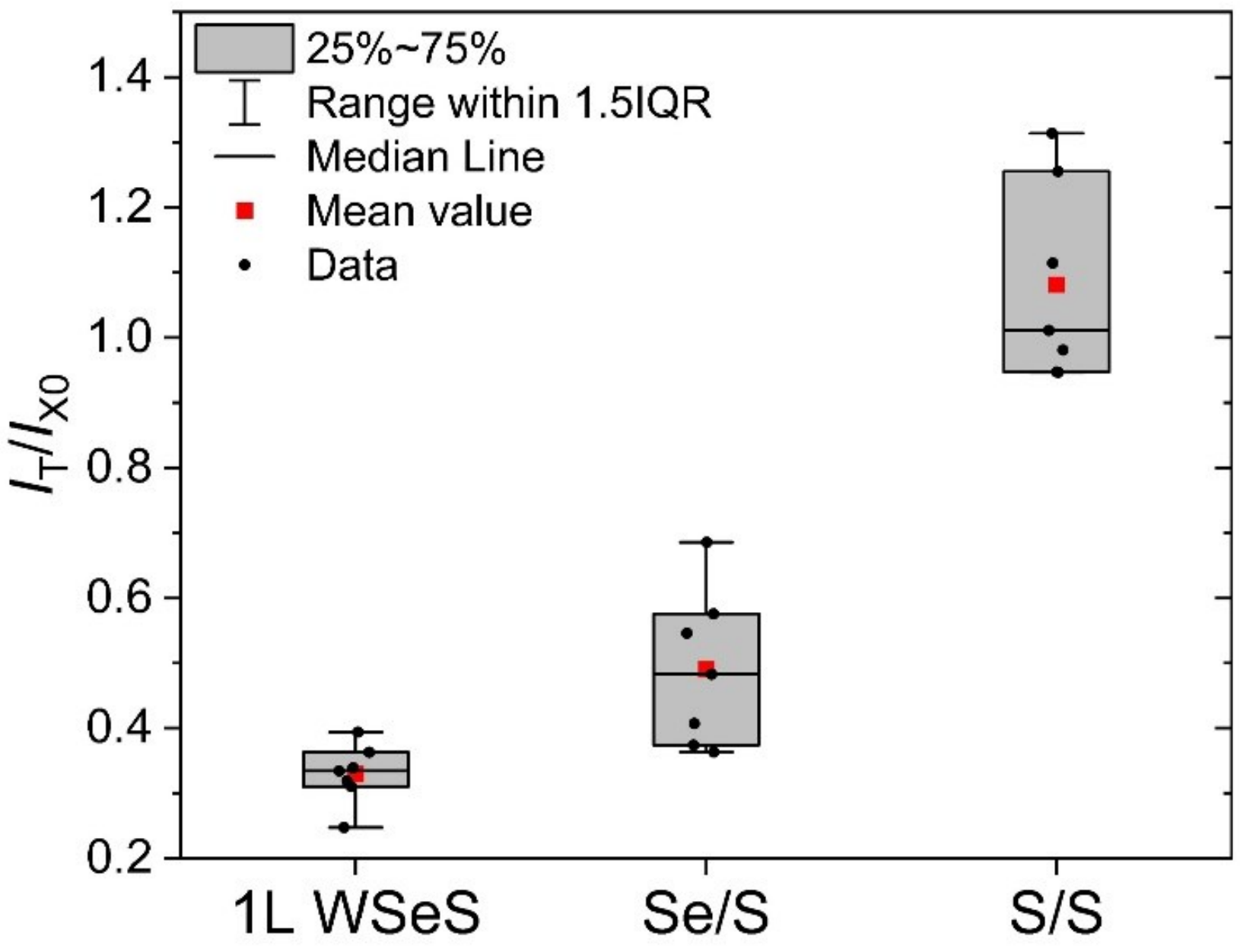

**Fig. S1**: Integrated intensity ratio of trion to exciton in monolayer (1L) WSeS, Se/S-, and S/S-ordered bilayer WSeS. The integrated intensity is extracted by deconvolving the Voigt function from the room-temperature PL spectra. Data points in the box chart come from different positions in each sample.

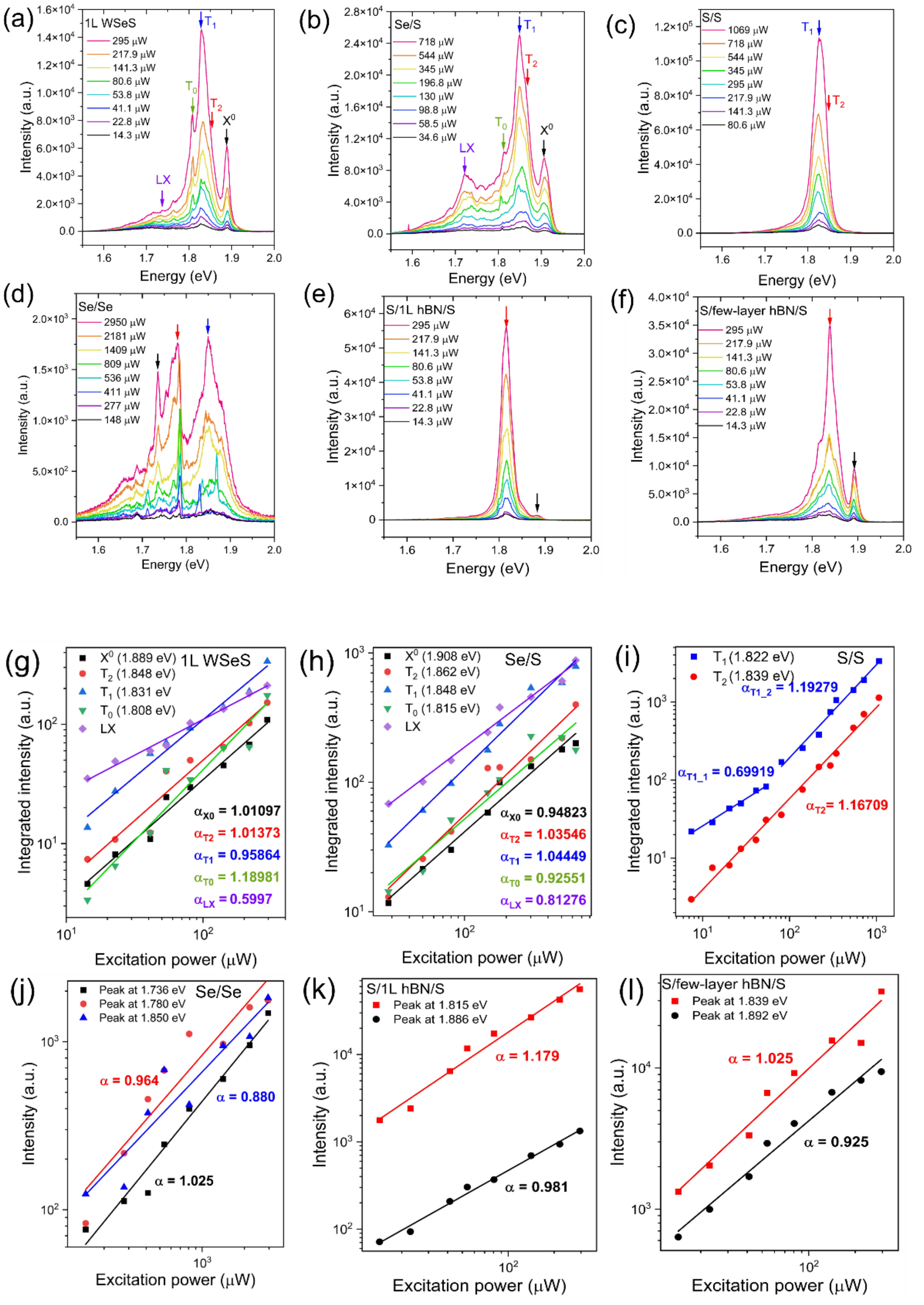

**Fig. S2**: Power-dependent PL obtained at 4.5 K. Spectra under different excitation powers in (a) monolayer WSeS, (b) Se/S-ordered homobilayer, (c) S/S-ordered homobilayer, (d) Se/Se-ordered homobilayer, (e) monolayer hBN-separated S/S-

ordered homobilayer, and (f) few-layer hBN-separated S/S-ordered homobilayer. Integrated intensities of each labelled peak from (a), (b), and (c) under different excitation powers. The integrated intensities are extracted by fitting the peaks with the Voigt function. Power-dependent absolute intensities of the labelled peaks from (d), (e), and (f).

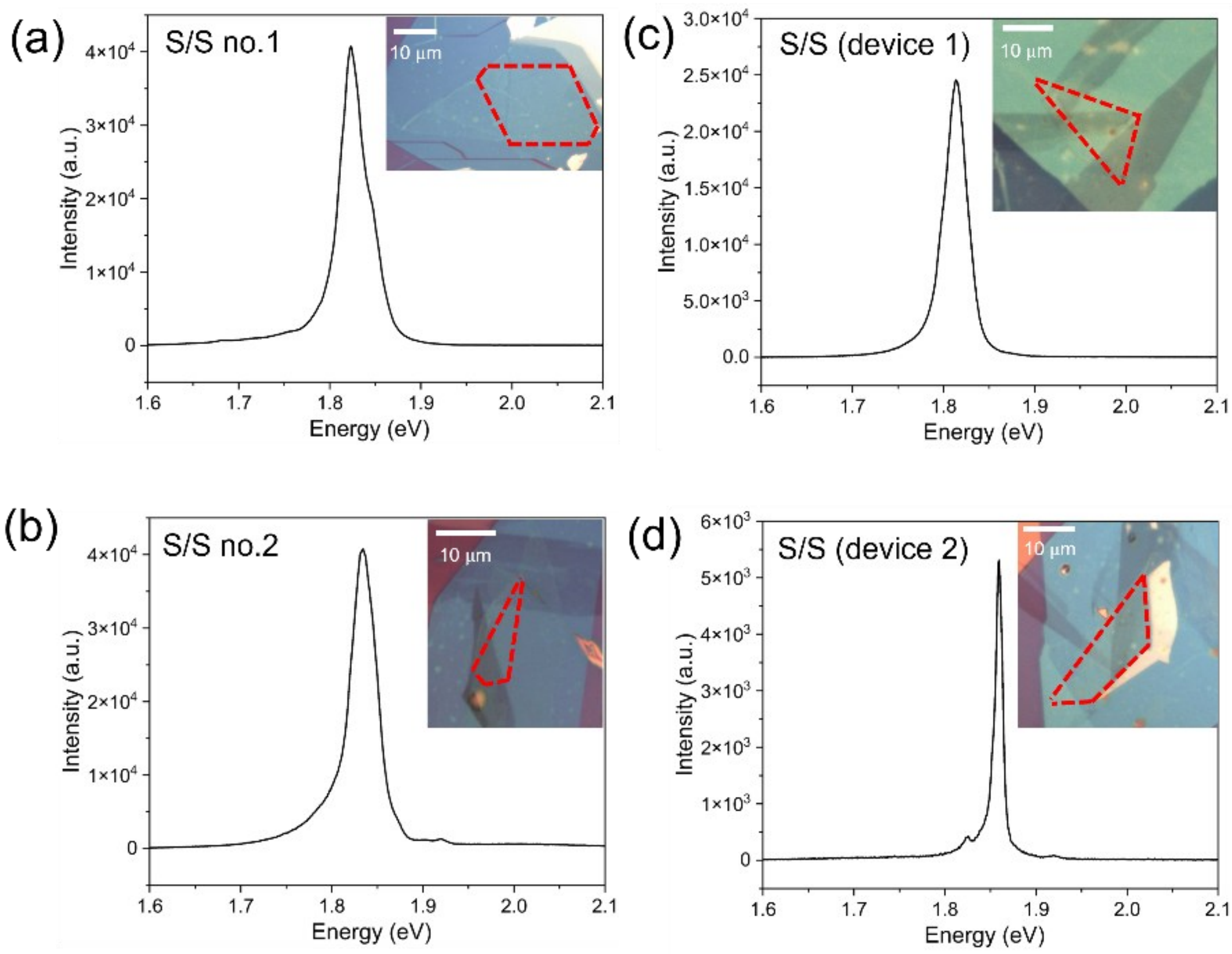

**Fig. S3**: PL spectra of four S/S-ordered homobilayer samples at 4.5 K. (a) and (b) The PL spectra of the samples shown in the main text and another hBN-encapsulated S/S bilayer, respectively. (c) and (d) The PL spectra of the device samples shown in Figure 2 and S4. Insets are the optical images of each sample. The red-dashed region is the S/S-ordered bilayer WSeS.

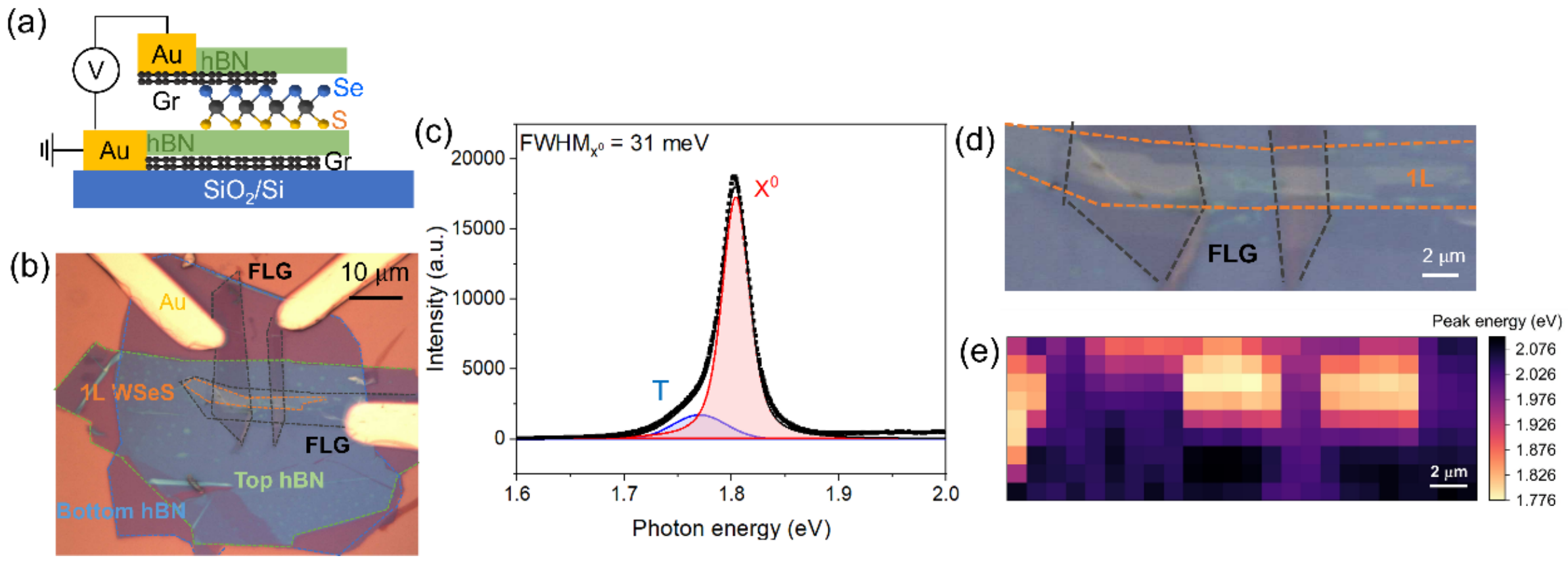

**Fig. S4**: (a) Schematic of the device of hBN-encapsulated monolayer WSeS, with few-layer graphene as electrodes. (b) Optical image microscopy of the device. (c) Room-temperature PL of the monolayer WSeS. (d) Zoomed region for obtaining PL mapping. (e) PL mapping of the resonance energy of the neutral exciton $X^0$ in the region of (d). The region contacted with graphene shows quenched PL, whereas the region without graphene shows obvious PL peaks at around 1.8 eV.

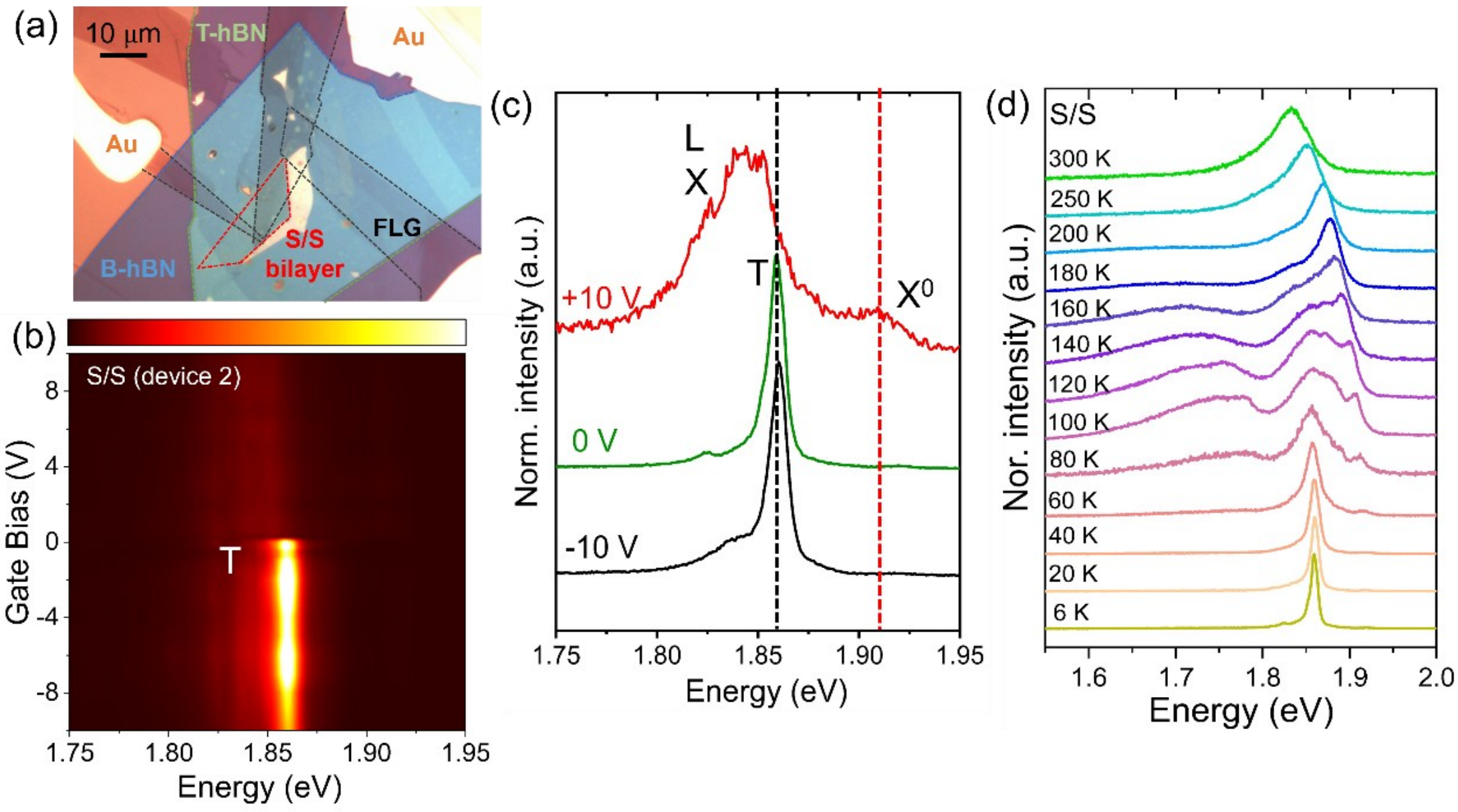

**Fig. S5**: (a) Optical image of the second device based on S/S-ordered WSeS homobilayer. (b) Color mapping of gate-dependent PL spectra of the second S/S device. (c) Corresponding PL spectra under several typical gate biases. (d) PL spectra without gate bias at different temperatures.

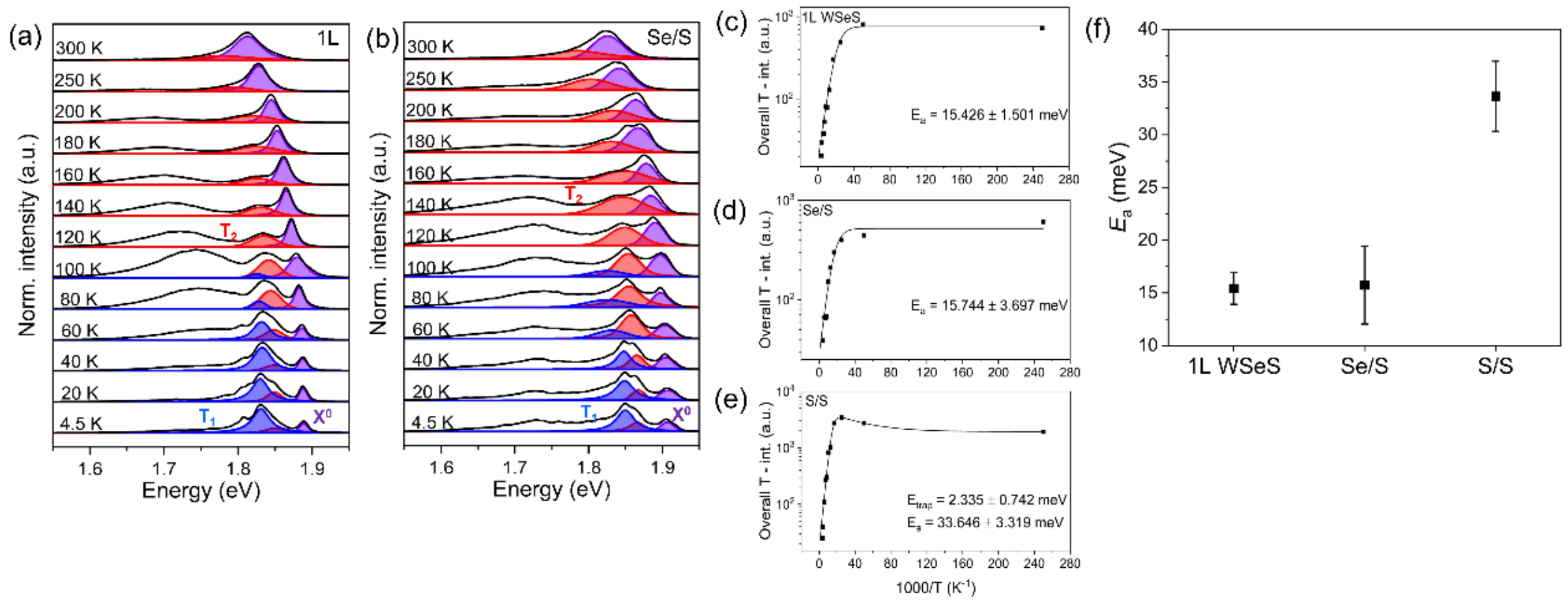

**Fig. S6**: Temperature dependence of the PL spectra of monolayer WSeS (a) and Se/S-ordered homobilayer (b). Temperature dependence of the integrated intensity of the overall trion ($T_1$ and $T_2$) of monolayer (c), Se/S-ordered (d), and S/S-ordered €homobilayer WSeS. (f) The activation energy for trion thermal dissociation $E_a$ in each sample extracted from the Arrhenius plot.

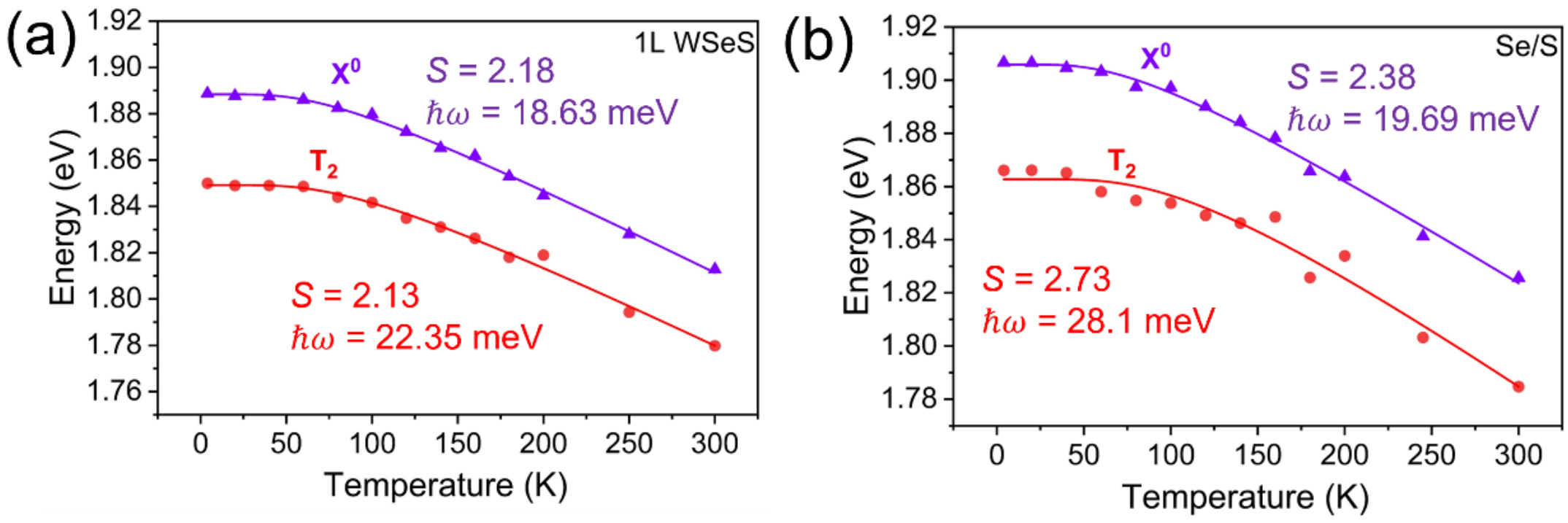

**Fig. S7**: Temperature dependence of the emission energies of the neutral exciton $X^0$ and standard trion $T_2$ of 1L WSeS (a) and Se/S-ordered bilayer WSeS. The electron-phonon coupling strength S and average phonon energy $\hbar\omega$ for each complex are extracted based on the O'Donnell model and indicated in each graph.

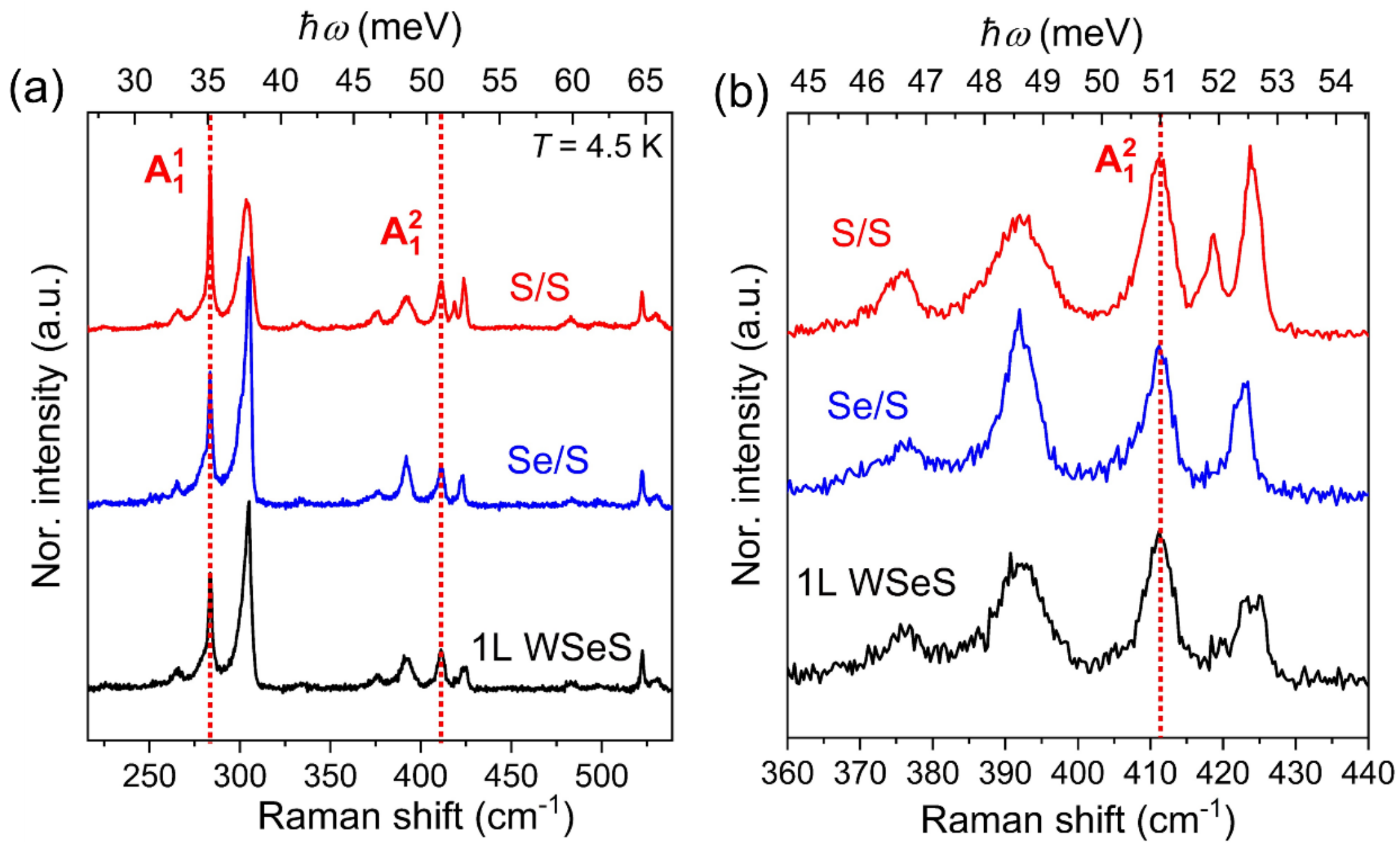

(c)

| | $A_1^2$ | | |
|---|---|---|---|
| | Wavenumber ($cm^{-1}$) | Height | FWHM ($cm^{-1}$) |
| 1L | 411.17767 | 0.84716 | 4.04429 |
| Se/S | 411.0851 | 0.82417 | 3.97358 |
| S/S | 411.03659 | 1.023 | 4.13117 |

**Fig. S8**: (a) Raman spectra of monolayer, Se/S-, and S/S-ordered homobilayer WSeS. (b) Corresponding Raman spectra in the range from 45 meV to 55 meV. The exact parameters of $A_1^2$ out-of-plane mode in each sample, including wavenumber, intensity, and full width at half maximum (FWHM), are listed in the table (c).

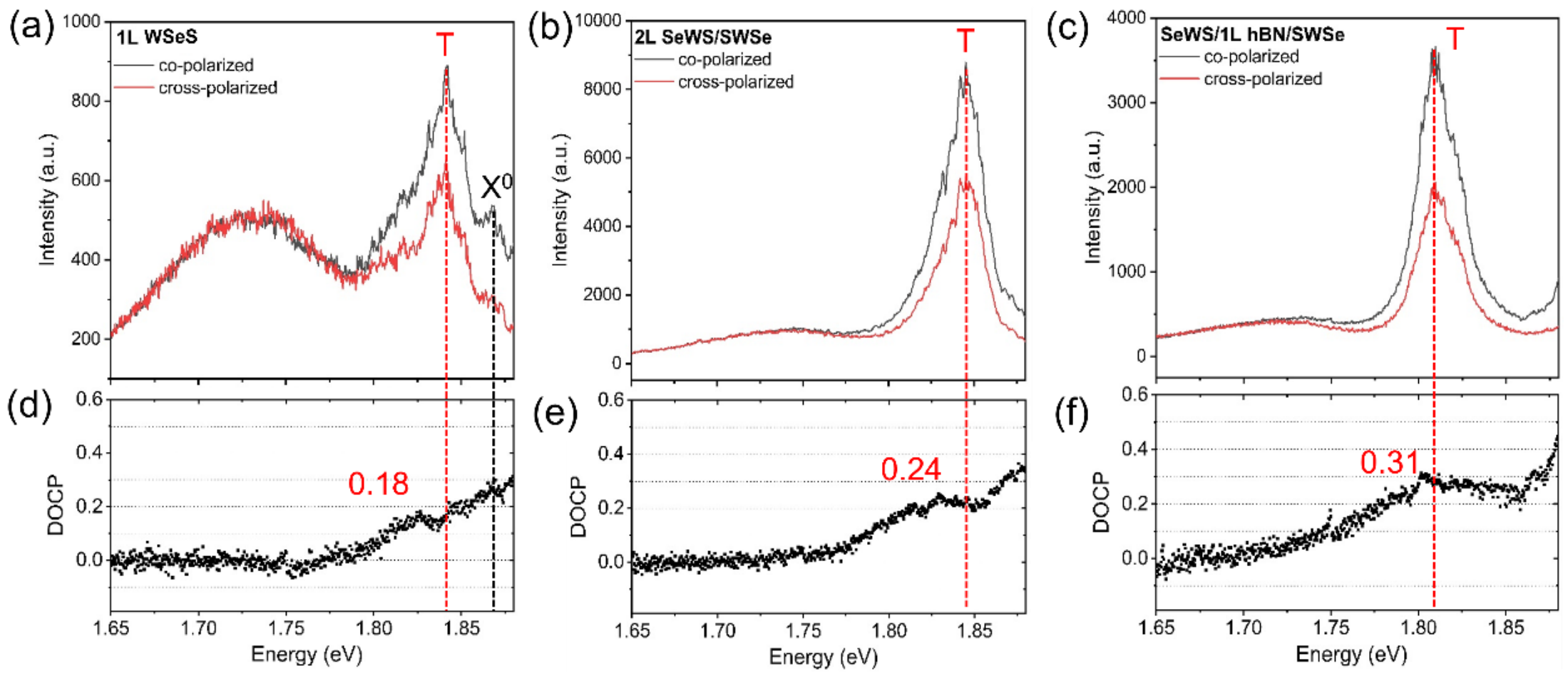

**Fig. S9**: (a)-(c) Circular polarization PL spectra of monolayer WSeS, S/S-ordered homobilayer, and monolayer hBN-separated S/S-ordered homobilayer. (d)-(f) Degree of circular polarization (DOCP) at different emission energies. DOCP of trion in each sample is listed in the graph.

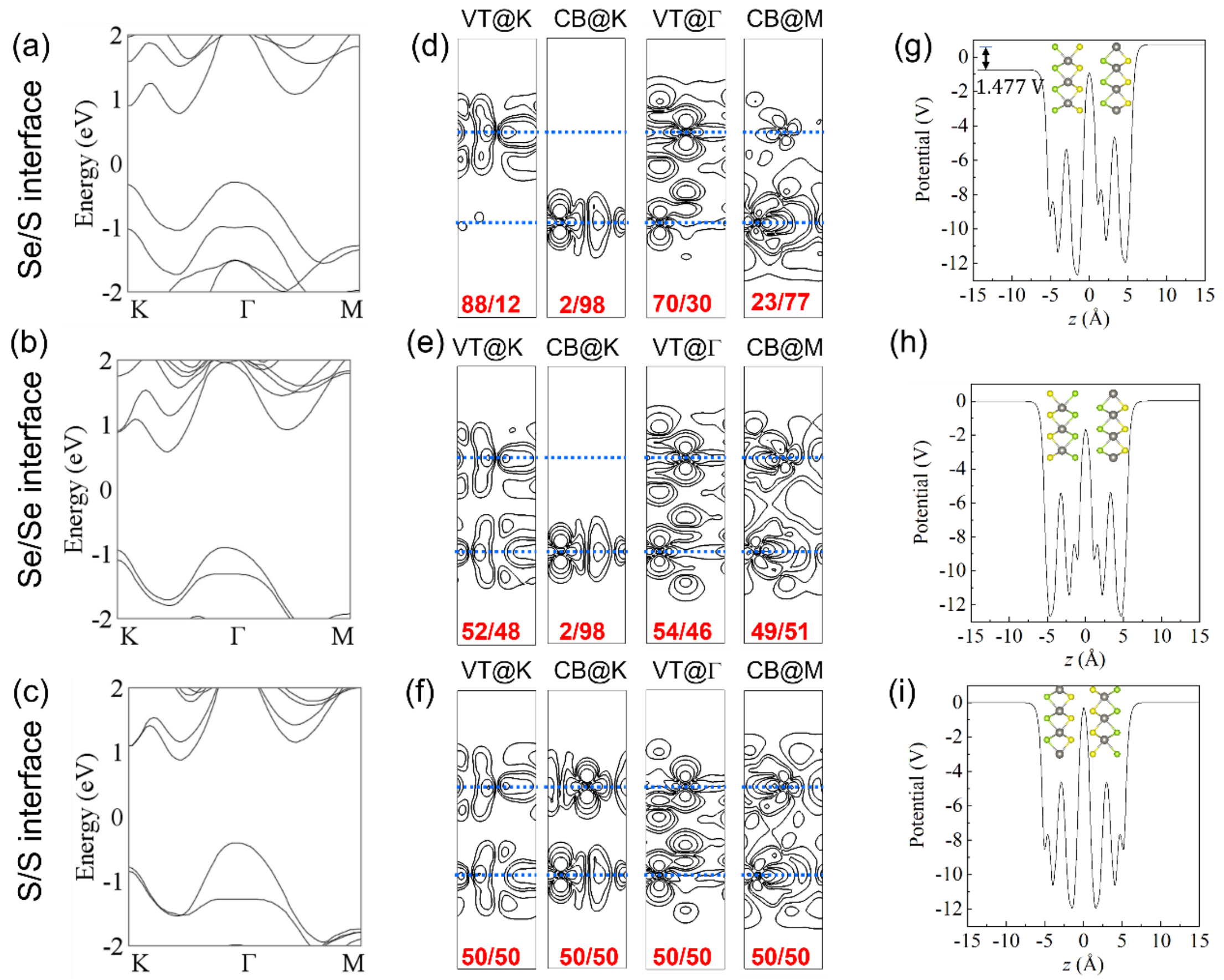

**Fig. S10**: Electronic band structures of Janus WSeS homobilayers with (a) Se/S, (b) Se/Se, and (c) S/S stacking orders. Contour plots of the squared wavefunction of the highest valence band and the lowest conduction band of Janus WSeS homobilayers with (d) Se/S, (e) Se/Se, and (f) S/S stacking orders. Each contour represents double/half of the adjacent contour. The numerical values represent the proportions of the wavefunction distributed over the upper and lower layers, respectively. Electrostatic potential of the Janus homobilayers with (g) Se/S, (h) Se/Se, and (i) S/S stacking orders.

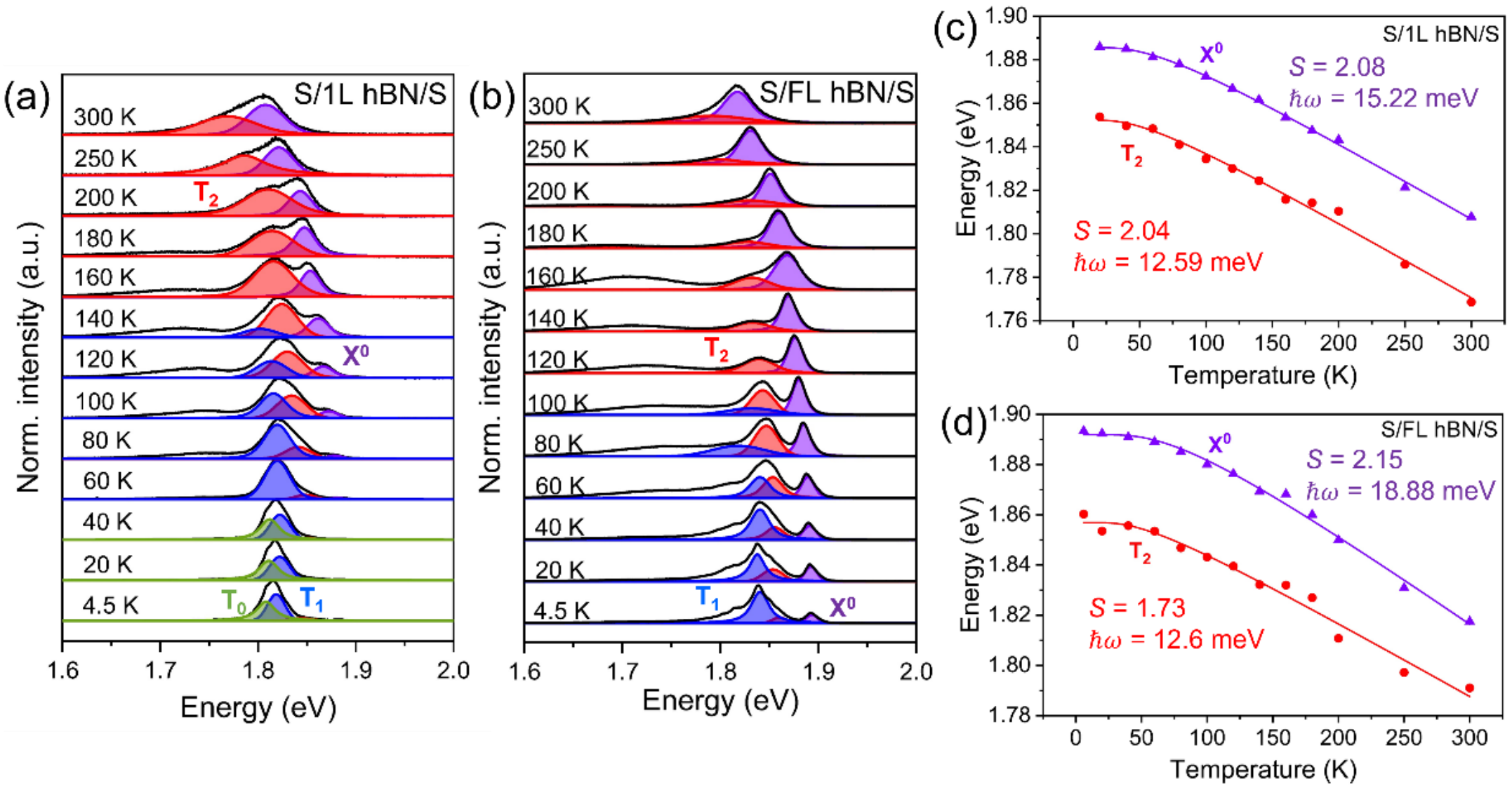

**Fig. S11**: Temperature evolution of PL spectra of monolayer hBN-separated (a) and few-layer hBN-separated (b) S/S bilayer WSeS. Temperature dependence of the emission energies of the neutral exciton $X^0$ and standard trion $T_2$ of the monolayer hBN-separated (c) and few-layer hBN-separated bilayer WSeS (d). The electron-phonon coupling strength S and average phonon energy $\hbar\omega$ for each complex are extracted based on the O'Donnell model and indicated in each graph.